\documentclass[twocolumn,preprintnumbers,amsmath,revsymb,nofootinbib]{revtex4-1}

\usepackage{amsfonts}
\usepackage{amssymb}
\usepackage{wasysym}
\usepackage{mathrsfs}
\usepackage{fontenc}
\usepackage{graphicx}
\usepackage{dcolumn}
\usepackage{bm}
\usepackage{feynmf}
\usepackage{verbatim}
\usepackage{hyperref}
\usepackage{subfigure}
\usepackage{enumerate}
\usepackage{slashed}
\usepackage{array}

\newcommand{\ds}{\displaystyle×}

\newcommand{\tev}{\, {\rm TeV}}
\newcommand{\gev}{\, {\rm GeV}}

\newcommand{\AFB}{A_{FB}}
\newcommand{\met}{\,/\hspace{-0.25cm}E_T}

\newcommand{\ie}{{i.e.}\ }
\newcommand{\beq}{\begin{equation}}
\newcommand{\eeq}{\end{equation}}

\newcommand{\thickhline}{%
    \noalign {\ifnum 0=`}\fi \hrule height 1pt
    \futurelet \reserved@a \@xhline
}

\begin{document}

\title{Tracking Down the Top Quark Forward-Backward Asymmetry with Monotops}

\author{Abhishek Kumar$^1$}
\email{abhishek@triumf.ca}
\author{John N. Ng$^1$}
\email{misery@triumf.ca}
\author{Andrew Spray$^2$}
\email{andrew.spray@coepp.org.au}
\author{Peter T. Winslow$^{1,3}$}
\email{pwinslow@triumf.ca}
\date{\today}

\affiliation{\vspace*{.1in}$^1$Theory Group, TRIUMF, 4004 Wesbrook Mall, Vancouver, BC, Canada, V6T 2A3}

\affiliation{\vspace*{.1in}$^2$ARC Centre of Excellence for Particle Physics at the Terascale, School of Physics, University of Melbourne, Victoria 3010, Australia}

\affiliation{\vspace*{.1in}$^3$Department of Physics and Astronomy, University of British Columbia, Vancouver, BC, V6T 1Z1, Canada}

\begin{abstract}
\indent \indent We revisit the possibility that the top quark forward-backward asymmetry arises from the on-shell production and decay of scalar top partners to $t\bar{t} + \slashed{E}_T$. Although the asymmetry is produced by t-channel exchange of a light mediator, the model remains unconstrained by low energy atomic parity violation tests. An interesting connection to the active neutrino sector through a Type-I seesaw operator helps to evade stringent monojet constraints and opens up a richer collider phenomenology. After performing a global fit to top data from both the Tevatron and the LHC, we obtain a viable region of parameter space consistent with all phenomenological and collider constraints. We also discuss the discovery potential of a predicted monotop signal and related lepton charge asymmetry at the LHC.
\end{abstract}

\maketitle

\section{Introduction}\label{Intro}

The large mass of the top quark has long been thought to imply a possible sensitivity to new physics (NP) near the electroweak scale. It is then extremely suggestive that the top quark forward-backward asymmetry ($A_{FB}^{t \bar{t}}$), as measured by both the CDF~\cite{Aaltonen:2011kc} and D$\O$~\cite{Abazov:2011rq} experiments, remains one of the few significant deviations from the predictions of the Standard Model (SM). Although both collaborations observe a preference for the production of $t$ quarks in the direction of the proton beam, only CDF observes a strong dependence on the invariant mass of the top quark pairs. This dependence has persisted as CDF has redone their analysis, using the full Tevatron data set of 9.4 fb$^{-1}$~\cite{Aaltonen:2012it}. Although the parton level asymmetry in the high invariant mass bin ($M_{t\bar{t}}>$ 450 GeV) has decreased in their more recent analysis, $A_{FB}^{t \bar{t}, high}$=$(29.5 \pm 6.5)\%$, it remains a $\sim$2.5$\sigma$ deviation from the current (NLO QCD including electroweak corrections) SM prediction of $(A_{FB}^{t \bar{t}, high})^{SM}$=$(12.9 \pm 0.7)\%$~\cite{Bernreuther:2012sx}. \\

Many NP models have been proposed to explain the asymmetry, each of which are characterized by either the exchange of new particles in the $s$- or $t$-channel of $t\bar{t}$ production or effective field theory methods (for a recent review see~\cite{Kamenik:2011wt} and references within). Recently, low energy precision tests of atomic parity violation (APV) were shown to strongly disfavor $t$-channel models with light mediators~\cite{Gresham:2012wc}. However, the case where $A_{FB}^{t \bar{t}}$ is generated from on-shell production and decay of scalar top partners to $t \bar{t} + \met$ (first mentioned in ref.~\cite{Isidori:2011dp}), although dependent on $t$-channel exchange to generate $A_{FB}^{t \bar{t}}$, is not constrained by APV. \\

In ref.~\cite{Isidori:2011dp}, a single, light, stable Majorana fermion is introduced and considered as a potential dark matter candidate. However, tension exists with this dark matter interpretation~\cite{Hektor:2011ms}. Furthermore, the parameter space of ref.~\cite{Isidori:2011dp} is constrained by current LHC bounds on monojets and jets+$\met$ signals, which arise in the model. In our model, we consider a triplet of unstable Majorana fermions and establish a possible connection to the active neutrino sector via a Type-I seesaw operator. While the mechanism for generating $A_{FB}^{t \bar{t}}$ is similar, the Majorana fermions in our model mix with active neutrinos, allowing them to decay to SM final states on detector length scales. These additional decay channels aid in evading tension with current LHC bounds from monojets and jets+$\met$ searches, rendering a viable parameter space that is consistent with top quark data and current LHC bounds. In addition to this, we also discuss search strategies and the LHC discovery potential for a monotop signal and an associated lepton charge asymmetry. \\


This paper is organized as follows: in Section II, we describe our model and its connection to the active neutrino sector. Section III describes our fit to the Tevatron and LHC top data and, in Section IV, we discuss phenomenological constraints from APV, Higgs searches, and flavor observables. In section V, we describe our analysis of the LHC signatures based on our fit, in particular the monotop signal and associated lepton charge asymmetry. Finally, in section VI, we present our final comments and conclusions. 

\section{Model}\label{sec:Model}

We add four new states to the SM, a color triplet scalar $\xi$ with the same gauge quantum numbers as the right-handed top quark, i.e., $\xi$:$(3, 1, 2/3)$ and three gauge-singlet Majorana fermions $\chi$=$(\chi_{u},\chi_{c},\chi_{t})$. The $\chi$ states correspond to right-handed neutrinos and so, consquently, $\xi$ is a leptoquark in our model. We emphasize here that the $\chi$ states are Majorana fermions and do not carry flavor quantum numbers, i.e., the subscripts only denote their couplings to the corresponding quark flavors\footnote{Recently, ref.~\cite{Kumar:2013hfa} proposed a model with similar field content, where the $\chi$'s form a triplet of Dirac fermions carrying flavor charges with their interactions constrained by Minimal Flavor Violation (MFV). The model explains $A_{FB}^{t \bar{t}}$ and also contains the lightest Dirac fermion, $\chi_{t}$, as a viable DM candidate.}. The Lagrangian is given by
\begin{eqnarray}
&\mathscr{L} = \mathscr{L}_{SM} + |D_\mu \xi|^2 + \sum_i \bar{\chi}_{i} i \slashed{\partial} \chi_i + \mathscr{L}_{mixing} - V(H,\xi) & \notag \\
&  - \lambda_{u} \overline{u}_{R} \xi \chi_{u}^{c} - \lambda_{c} \overline{c}_{R} \xi \chi_{c}^{c} - \lambda_{t} \overline{t}_{R} \xi \chi_{t}^{c} +  \text{h.c.} &
\label{ModelLag}
\end{eqnarray}
where $\mathscr{L}_{SM}$ is the SM Lagrangian, $i$=($u,c,t$), $\mathscr{L}_{mixing}$ mixes the $\chi$ states with the active neutrino sector, and $H$ is the SM Higgs doublet. The scalar potential, $ V(H,\xi)$, which couples the Higgs to $\xi$, does not contribute to $A_{FB}^{t \bar{t}}$ and will be discussed in Section~\ref{sec:Higgs}.

\begin{figure}[t]
\includegraphics[scale=.5]{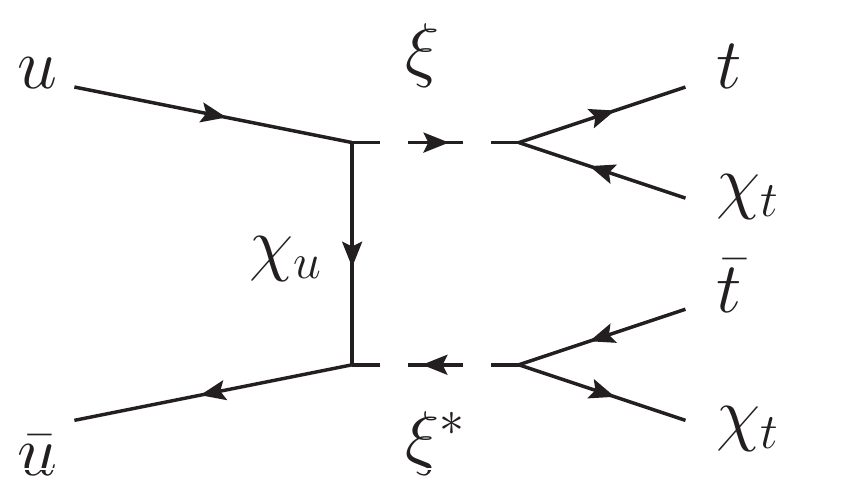}
\caption{\textit{NP production of the top quark forward-backward asymmetry through on-shell production and decay of scalar top partners.}}
\label{fig:AFB_process}
\end{figure}

The top quark forward-backward asymmetry is generated in the on-shell production of $\xi\xi^{*}$, which subsequently decays to $t\bar{t}\chi_{t}\chi_{t}$, as shown in FIG.~\ref{fig:AFB_process}. Requiring small mass splittings between $\xi$ and the top quark limits the amount of $\slashed{E}_T$ from $\chi_t$, simulating a $t \bar{t}$ final state. This process does not interfere with SM $t\bar{t}$ production and generates a large NP asymmetry through a Rutherford enhancement from the t-channel exchange of $\chi_u$. Furthermore, $\xi \xi^{*}$ production is $p$-wave suppressed, significantly reducing the NP contribution to the total $t\bar{t}$ production cross-section. A large $\lambda_{t}$ coupling increases the branching ratio, BR($\xi\to t \chi_{t}$), enhancing the population of top quarks in the final state and amplifying $A_{FB}^{t \bar{t}}$. Simultaneously, a small $\lambda_{u}$ coupling decreases BR($\xi\to u\chi_{u}$), thereby enhancing BR($\xi\to t \chi_{t}$) while also reducing the NP production cross-section. However, if $\lambda_u$ is too small, $A_{FB}^{t \bar{t}}$ becomes too suppressed. Consequently, agreement with the Tevatron data requires large $\lambda_{t}$ and $\mathcal{O}$(1) $\lambda_{u}$. \\

Our model also has an interesting connection with the active neutrino sector. Mixing between the active and sterile neutrinos via type-I seesaw operators, $\kappa_{i j} \bar{L}_{i} \tilde H \chi_{j}$, readily occurs without any $\mathcal{Z}_2$ symmetry to forbid it. These interactions mix the sterile and active neutrinos via a 6$\times$6 mass matrix which can be organized into a 3$\times$3 block form as
\begin{eqnarray}
&- \mathscr{L}_{mixing} = \frac{1}{2} (M_\chi)_{i} \overline{\chi^c_i} \chi_i + \kappa_{ij} \overline{L}_i \tilde{H} \chi_j + h.c.& \notag \\
& = \frac{1}{2} 
\left(
\begin{array}{cc}
\overline{\nu} & \overline{\chi^c} 
\end{array}
\right)
\left(
\begin{array}{cc}
0 & M_D \\
M_D^T & M_\chi
\end{array}
\right)
\left(
\begin{array}{c}
\nu^c \\
\chi
\end{array}
\right) + h.c.
\label{Mixing}
\end{eqnarray}
where $\nu$ and $\chi$ represent active and sterile neutrinos, $\tilde{H} = i \sigma_2 H^*$, $(M_D)_{ij} = \kappa_{ij} \; v / \sqrt{2}$, and $v$ is the Higgs vev. $M_\chi$ is the singlet mass matrix, which we take to be diagonal. The 6$\times$6 symmetric mass matrix can be diagonalized by a unitary matrix $\mathscr{U}$ as 
\begin{eqnarray}
\mathscr{U}^\dagger 
\left(
\begin{array}{cc}
0 & M_D \\
M_D^T & M_\chi
\end{array}
\right) \mathscr{U}^* = 
\left(
\begin{array}{cc}
m_\nu & 0 \\
0 & M_\chi
\end{array}
\right)
\end{eqnarray}
where, at leading order, $\mathscr{U}$ can be parameterized as~\cite{Asaka:2011pb} 
\begin{eqnarray}
\mathscr{U} = \left(
\begin{array}{cc}
U & \Theta \\
- \Theta^\dagger U & 1
\end{array}
\right).
\end{eqnarray}
Here, $U$ is the 3$\times$3 PMNS matrix that diagonalizes the active neutrino mass matrix from the seesaw mechanism, i.e., $U^\dagger ( - M_D M_\chi^{-1} M_D^T ) U^*$ = diag$(m_{\nu_1}, m_{\nu_2}, m_{\nu_3})$ and $\Theta$ is the 3$\times$3 matrix mixing the active and singlet neutrinos, given by $\Theta = M_D M_\chi^{-1}$ at leading order. 
In light of bounds on the energy density of active neutrinos from the WMAP~\cite{Spergel:2003cb} and Planck~\cite{Ade:2013zuv} collaborations, we take $m_{\nu_i} \sim 0.5 \times 10^{-9}~\gev$ and obtain $\kappa \sim \sqrt{m_{\nu_{i}} m_{\chi}/v^{2}} \sim \sqrt{ 10^{-14} \; (m_\chi / \textrm{GeV})}$. From this, we estimate the active-sterile mixing,
$\Theta \sim M_{D}/M_{\chi} \sim \sqrt{m_{\nu}/m_{\chi}} \sim \sqrt{0.5 \times 10^{-9} \; (\textrm{GeV} / m_{\chi})}$. Taking $\mathcal{O}(10~\gev)$ sterile neutrino masses, we can treat the mixing perturbatively in the mass basis,
\begin{eqnarray}
&- \mathscr{L}_{mixing}  = \frac{1}{2} m_{\nu_i} \overline{\nu_i}  \nu_i^c + \frac{1}{2} M_{\chi_{j}} \overline{\chi_{j}^{c}}  \chi_j & \notag \\
& + \Theta_{ij} m_{\chi_j} \overline{\nu_i} \chi_j & \end{eqnarray}
where $i=1,2,3$ and $j=u,c,t.$ \\

\begin{figure}[t]
\includegraphics[scale=.65]{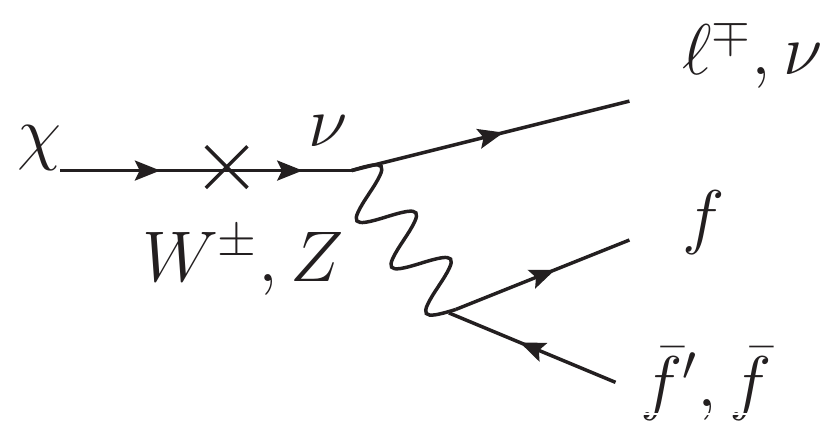}
\caption{\textit{The dominant decay channels for $\chi$ states.}}
\label{fig:chi_decay}
\end{figure}

\begin{figure}[h!]
\includegraphics[scale=.625]{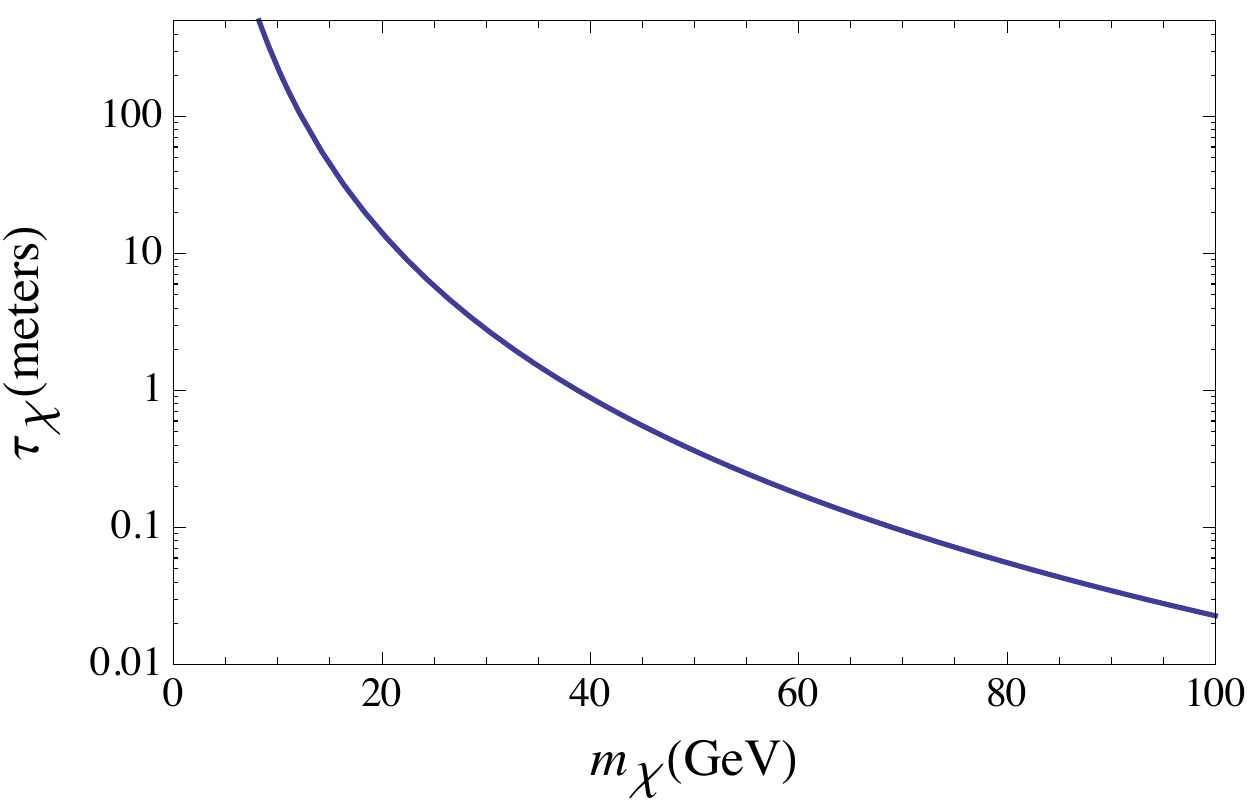}
\caption{\textit{The decay length of a given $\chi$ state as a function of its mass in meters.}}
\label{fig:chiDecay}
\end{figure}

Mixing with the active neutrinos opens new decay channels for the $\chi$ states. After mixing, a $\chi$ state can undergo a 3-body weak decay through both neutral and charged current-current interactions, as shown in FIG.~\ref{fig:chi_decay}. Accounting for all possible decay channels imparts a decay length to a given $\chi$ state that is particularly sensitive to its mass, i.e., 
\begin{eqnarray}
\tau_\chi = \gamma / \Gamma_\chi \simeq \frac{10^6}{ m_\chi^4 } \text{ m} ,
\label{eq:chi_width}
\end{eqnarray}
where the boost factor, $\gamma$, has been included. We have studied the boost factor in all processes that we consider and calculate an average of $\gamma \simeq 5$. 
Using this value in Eq.~(\ref{eq:chi_width}) yields the result, which is shown in FIG.~\ref{fig:chiDecay}. \\

The decay length peaks sharply at low mass, allowing for the possibility of light $\chi$ states remaining stable on collider length scales and simply registering as $\slashed{E}_T$. However, if a $\chi$ state is sufficiently massive, it will decay within both the ATLAS and CMS detectors, potentially leaving charge tracks. In particular, for $m_{\chi}\gtrsim 40~\gev$, the decay length is less than 1 meter (which we take as the distance to the ATLAS and CMS calorimeters). From a phenomenological point of view, we choose $\{m_{\chi_u}, m_{\chi_c}, m_{\chi_t}\} = \{45\gev, 45\gev, 10\gev\}$,  such that $\chi_{u}$ and $\chi_c$ decay inside the calorimeters while $\chi_{t}$ decays far outside the detectors. As a result, $\chi_{t}$ is stable on detector length scales, registering as $\slashed{E}_T$, so that $A_{FB}^{t \bar{t}}$ is still generated in the NP production of $t\bar{t}+\met$. Furthermore, as will be discussed in section~\ref{sec:LHC}, the instability of $\chi_u$ and $\chi_c$ on detector length scales is essential to evading tight LHC constraints from monojet searches.

\section{Global Fit to Top Data from Tevatron and LHC}\label{sec:GlobalFit}

The top quark forward-backward asymmetry can be factorized into the SM and NP contributions as
\begin{eqnarray}
 \AFB =  \AFB^{SM} \left( 1 + \frac{\sigma^{NP}}{\sigma^{SM}} \right)^{-1} + \AFB^{NP} \left( 1 + \frac{\sigma^{SM}}{\sigma^{NP}} \right)^{-1}
 \label{AFB_factorization}
\end{eqnarray} 
where $\AFB^{SM}$ is the SM asymmetry, generated at $\mathcal{O}(\alpha_{s}^{3})$ in the cross-section, and $\AFB^{NP}$ is the NP asymmetry, generated at tree-level. In our analysis, we calculate NP effects at tree-level only and assume the same K-factors for SM and NP $t\bar{t}$ production. The charge asymmetry at the LHC is similarly defined.

Our model is described by 3 parameters, $\lambda_u$, $\lambda_t$, and $m_\xi$ (we set $\lambda_c=0$ as discussed in Section~\ref{sec:Flavor}). To determine the viable regions of parameter space, we perform a global $\chi^2$ fit of the model parameters to the cross section and asymmetry data from both the Tevatron and LHC. \\

 In our fit, we consider the following:    
\begin{itemize}   
\item[-] CDF and D$\O$ combination of inclusive cross section measurements, $\sigma_{t \bar{t}} = 7.65 \pm 0.41$ (stat+syst) pb~\cite{TevEWworkingGroup}, along with the NNLO+NNLL SM calculation,  $\sigma_{t \bar{t}} = 7.067^{+0.143}_{-0.232}$(scale)$^{+0.186}_{-0.122}$(pdf) pb~\cite{Barnreuther:2012ws}. 
\item[-] Inclusive forward-backward asymmetry measurements from both CDF, $A_{FB}^{t \bar{t}}= 0.164 \pm 0.047$ (stat+syst)~\cite{Aaltonen:2012it}, and D$\O$, $A_{FB}^{t \bar{t}}= 0.196 \pm 0.065$ (stat+syst)~\cite{Abazov:2011rq}, along with the theoretical NLO SM calculation, $A_{FB}^{t \bar{t}} = 0.088 \pm 0.006$ (scale+pdf)~\cite{Bernreuther:2012sx}.  
\item[-] Differential cross section measurements from CDF~\cite{Aaltonen:2009iz} along with the theoretical NNLO SM calculation~\cite{Ahrens:2010zv}.
\item[-] Differential forward-backward asymmetry measurements from CDF~\cite{Aaltonen:2012it} along with the theoretical NNLO SM calculation which we take from~\cite{Aaltonen:2012it}.  
\item[-] Inclusive cross section measurements from ATLAS, $\sigma_{t \bar{t}} = 165 \pm 17$ (stat+syst) pb~\cite{ATLAS:2012gpa}, and CMS, $\sigma_{t \bar{t}} = 161.9 \pm 6.6$ (stat+syst)~\cite{Chatrchyan:2012bra}, along with the approximate NNLO SM calculation, $\sigma_{t \bar{t}} = 163 \pm 11$ (scale+pdf) pb~\cite{Kidonakis:2010bb}, for 7 TeV. 
\item[-] Inclusive charge asymmetry measurements from ATLAS, $A_C = -0.019 \pm 0.037$ (stat+syst)~\cite{ATLAS:2012an}, and CMS, $A_C = -0.013 \pm 0.04$ (stat+syst)~\cite{Chatrchyan:2011hk}, along with the SM approximate NLO calculation from MC@NLO, $A_C = 0.006 \pm 0.002$ (scale+pdf)~\cite{ATLAS:2012an}.  
\item[-] Differential charge asymmetry measurements from CMS~\cite{Chatrchyan:2012cxa} along with the SM approximate NLO calculation from POWHEG~\cite{Chatrchyan:2012cxa}.   
\end{itemize} 
To calculate the asymmetry, we simulate the NP process $pp \to \xi \xi^{*} \to t\bar{t}\chi_{t}\chi_{t}$ at $\sqrt{s}= 1.96$ \& $7$ TeV at the Tevatron and LHC respectively, using MadGraph5v1.5.10~\cite{Alwall:2011uj} and the CTEQ6L1 pdf set~\cite{Pumplin:2002vw}. 
\begin{figure*}[t!]
\centering
\mbox{
\subfigure{
\includegraphics[scale=.625]{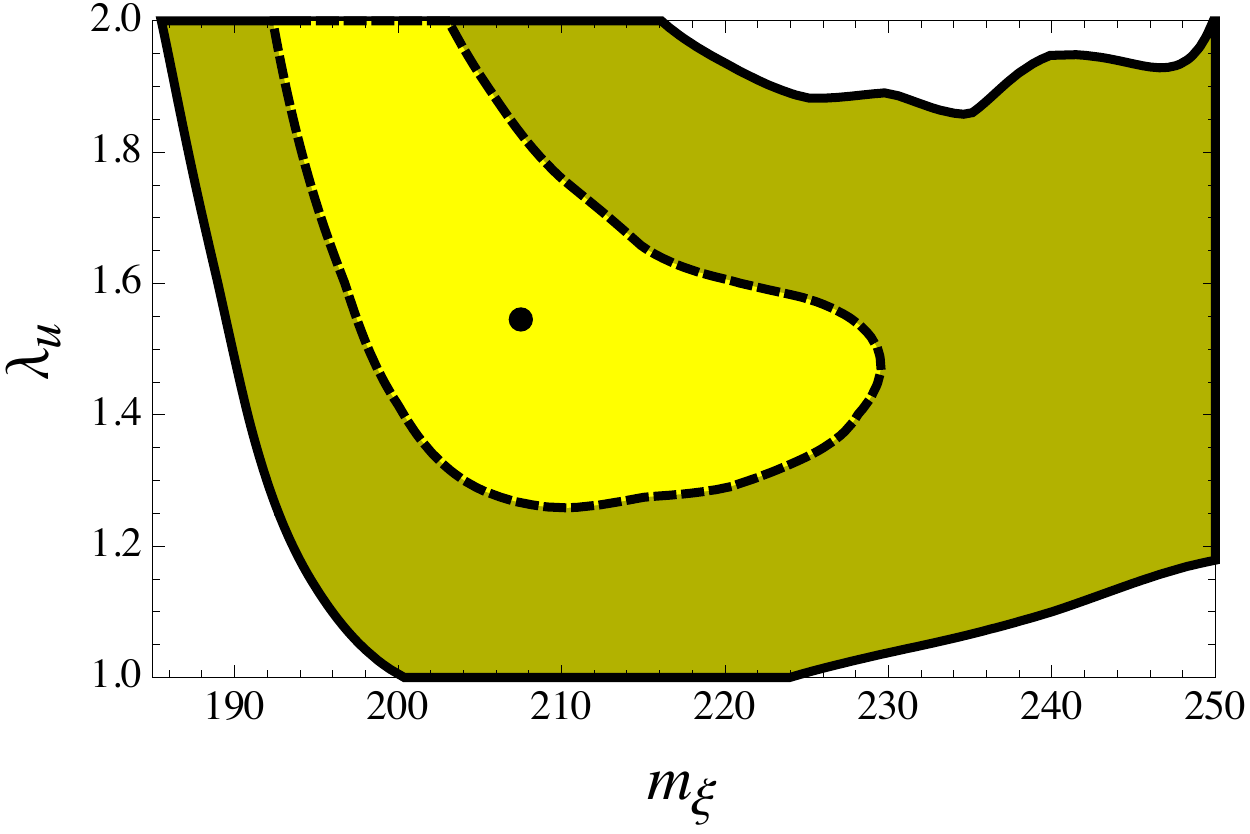}
\label{fig:yu_m}
}
\quad \quad \quad \subfigure{
\includegraphics[scale=.625]{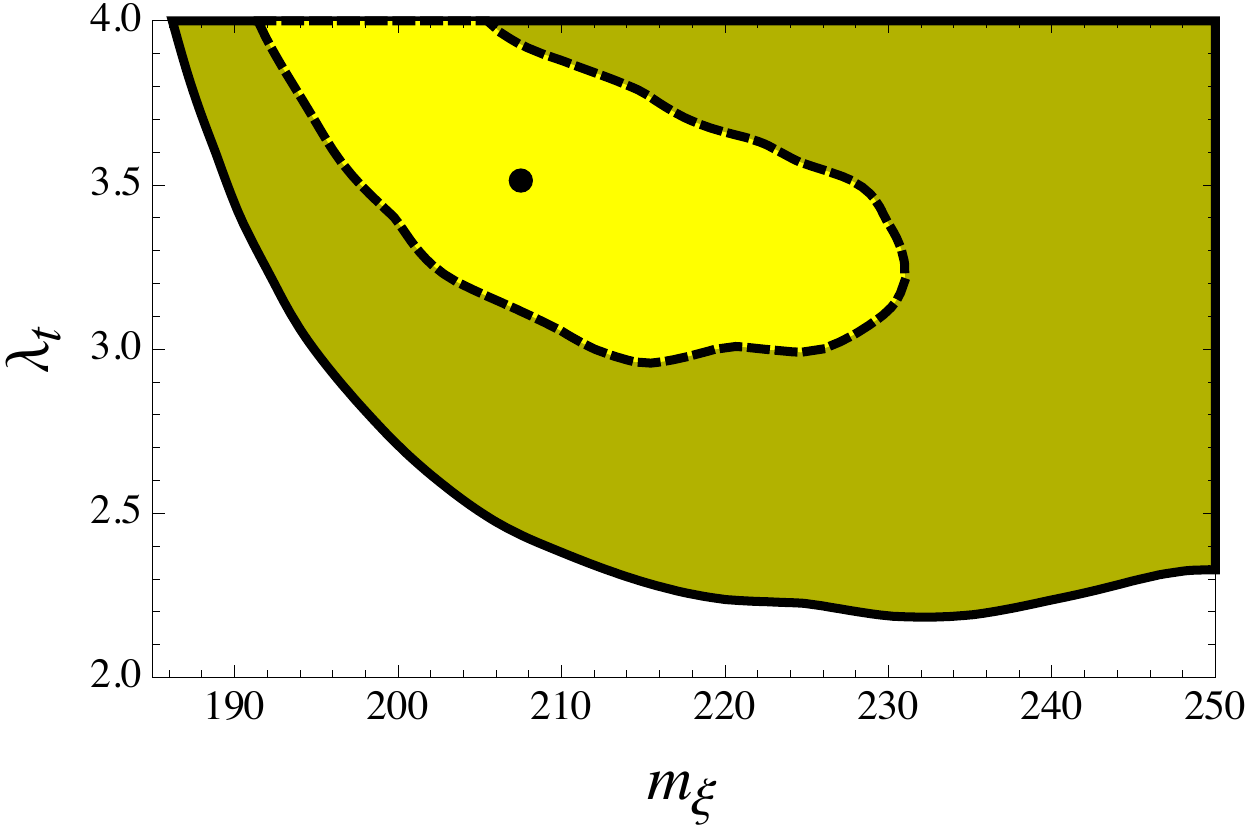}
\label{fig:yt_m}
}
}
\caption{\textit{Best fit regions in the $\lambda_u$-$m_\xi$ (left) and $\lambda_t$-$m_\xi$ (right) planes in a global $\chi^2$ fit to data from the Tevatron and the LHC. The light (dark) region is the 1$\sigma$ (2$\sigma$) region corresponding to $\Delta \chi^2 = $1(4). In each plot, the coupling that is not featured is fixed to its best fit value.}}
\label{fig:BestFitRegions}
\end{figure*}

To determine the region allowed by the current data, we construct a $\chi^2$ function of the three model parameters. At the best fit point, given by $\lambda_u$=1.55, $\lambda_t$=3.51, and $m_\xi$=207.6 GeV, we obtain a good fit to the data, i.e., $\chi_{min}^2/d.o.f$=0.98. The resulting best fit regions of parameter space are shown in the $\lambda_u$-$m_\xi$ and $\lambda_t$-$m_\xi$ planes presented in the left and right panels, respectively, of FIG.~\ref{fig:BestFitRegions}. The light (dark) yellow region corresponds to $\Delta \chi^2 = \chi^2 - \chi_{min}^2 = $1(4), which we refer to as the 1$\sigma$ (2$\sigma$) region. In both plots, the coupling that is not featured is fixed to its best fit value. 

\section{Phenomenological Constraints}\label{sec:Constraints}

In this section we explore the phenomenological constraints from atomic parity violation, Higgs searches, and flavor observables.

\subsection{Atomic parity violation}\label{sec:APV}

Parity violating electron-quark interactions, arising from $Z$ exchange, are described by current-current interactions below the weak scale,
\begin{eqnarray}
\mathscr{L}_{eq} = \frac{G_F}{\sqrt{2}} \sum_{u,d} \left( C_{1q} \overline{e} \gamma^\mu \gamma^5 e \overline{q} \gamma_\mu q + C_{2q} \overline{e} \gamma^\mu e \overline{q} \gamma_\mu \gamma^5 q \right) .
\end{eqnarray}
Only the first term contributes to APV in the limit of low momentum transfer, and at leading order, the SM values for the coefficients are $C_{1u}^{SM} = -\frac{1}{2} + \frac{4}{3} s_W^2$ and $C_{1d}^{SM} = \frac{1}{2} - \frac{2}{3} s_W^2$, where $s_W^2 \equiv \sin^2 \theta_W$. At the atomic level, this term leads to mixing between different energy levels with opposite parities, generating the nuclear weak charge
\begin{eqnarray}
Q_{wk} (N,Z) = - 2 \left( (2 Z + N) C_{1u} + (2N + Z) C_{1d} \right) .
\end{eqnarray}
Including higher order SM corrections to $C_{1q}$ leads to stringent constraints on NP models~\cite{Marciano:1982mm,*Marciano:1980pb}.

NP models at the weak scale can modify $C_{1q}$ by generating anomalous couplings between the $Z$ and the light quarks, 
\begin{eqnarray}
\mathscr{L}_{NP} = - \frac{g_2}{c_W} Z^\mu \left( a_{q,L}^{NP} \overline{q}_L \gamma_\mu q_L + a_{q,R}^{NP} \overline{q}_R \gamma_\mu q_R \right),
\end{eqnarray}
giving $C_{1q} = C_{1q}^{SM} + a_{q,L}^{NP} + a_{q,R}^{NP}$. Several models explaining the top quark forward-backward asymmetry with light $t$-channel mediators connecting $u$ and $t$ quarks are strongly disfavoured by APV constraints~\cite{Gresham:2012wc}. This is due to a sensitivity of $a_{q,R}^{NP}$ to electroweak symmetry breaking, which guarantees that it is proportional to the largest electroweak symmetry breaking generated mass available in the loop\footnote{The vertex correction and wave-function renormalization diagrams both contribute to $a_{q,R}^{NP}$ and would exactly cancel in the absence of electroweak symmetry breaking.}.

\begin{figure}[t!]
\includegraphics[scale=.6]{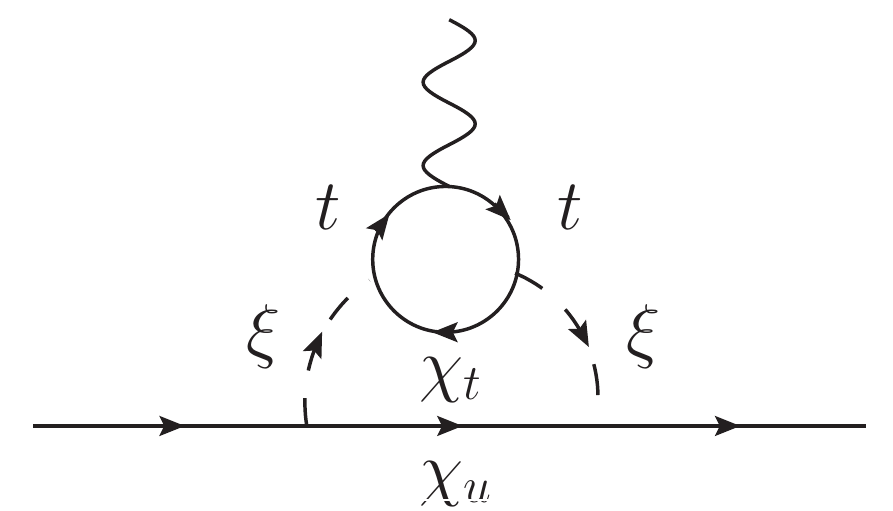}
\caption{\textit{The dominant 2-loop diagram for anomalous Zqq couplings contributing to APV.}}
\label{fig:APV_Diagrams}
\end{figure}

In our model, the 1-loop correction to $a_{u,R}^{NP}$ involves the $u$-quark only and is therefore negligible. However, at the 2-loop level, the $t$ quark enters (FIG~\ref{fig:APV_Diagrams}), leading to a correction which we estimate as
\begin{eqnarray}
a_{u,R}^{NP} \approx \frac{2 s_W^2 |\tilde{y}_u|^2 |\tilde{y}_t|^2}{3 (4 \pi)^4} \frac{m_t^2}{m_\xi^2} .
\end{eqnarray}
At the best fit point, our model predicts a correction $a_{u,R}^{NP} \sim 1 \times 10^{-4}$, an order of magnitude below the expected sensitivity of the upcoming proton weak charge measurement by the Qweak collaboration~\cite{VanOers:2007zz}. The reason for the decreased sensitivity in our model is the lack of a direct coupling between the $u$ and $t$ quarks.

\subsection{Higgs Phenomenology}\label{sec:Higgs}

The most general, renormalizable scalar potential coupling the top partner $\xi$ to the Higgs	field is
\begin{eqnarray}
&V(H, \xi) = - \mu^2 | H |^2 + \lambda ( |H|^2)^2 + g_\xi |\xi|^2 |H|^2 & \notag \\
& + \lambda_\xi |\xi|^4 + M_\xi^2 |\xi|^2 .&
\end{eqnarray}
Expanding about the minimum of the potential, while ensuring no color breaking minima, gives the following interactions
\begin{eqnarray}
&V(h, \xi) = \displaystyle \frac{\lambda}{2} ( h^2 + 2 v h )^2 + \frac{g_\xi}{2} |\xi|^2 ( h^2 + 2 v h ) & \notag \\
& + \lambda_\xi ( |\xi|^2 )^2 + m_\xi^2 |\xi|^2&
\label{Higgs_Potential}
\end{eqnarray}
where $m_\xi = \ds \sqrt{M_\xi^2 + g_\xi \frac{v^2}{2}}$ with $v$=246 GeV. These interactions can lead to significant effects in the production and decay of the Higgs. As the current Higgs data is consistent with the SM interpretation, a comparison with the data provides non-trivial constraints on $m_\xi$ and the Higgs portal coupling, $g_\xi$.

We assume that the 125 GeV resonance observed at the LHC~\cite{Aad:2012tfa,Chatrchyan:2012ufa} is due to the SM Higgs boson. The dominant effects of our model are then through $\xi$ loops which modify Higgs production through gluon-gluon fusion and Higgs to di-photon decays. Furthermore, an invisible decay channel, $h \to \chi_t \chi_t$ via $\xi$ loops, also modifies the Higgs width. However, the rate for this decay is generically $\mathcal{O}(10^{-2})$ relative to the $h \rightarrow b\bar{b}$ rate for the entire region of parameter space of interest.  \\

In order to connect with the data, we consider the signal rates for the various decay channels, defined as 
\begin{eqnarray}
\mu_{XX} = \frac{\sum_i \epsilon_i \sigma(i \rightarrow h)}{\sum_i \epsilon_i \sigma(i \rightarrow h)^{SM}} \frac{\text{BR} ( h \rightarrow XX )}{\text{BR} ( h \rightarrow XX )_{SM}}
\label{signal_rates}
\end{eqnarray}
where we have summed over the different production modes, $i$, and included efficiency factors, $\epsilon_i$, that are related to the acceptance for a given choice of cuts. 
\begin{table}[t!]
\centering
\begin{tabular}{|c|c|c|c|}
\hline 
channel & $\mu$ & (ggF, VBF+VH) & Ref. \\
\hline
$\gamma \gamma$ incl. & 1.6 $\pm$ 0.3 & (88\%,12\%) & \cite{ATLAS:2013gg} \\
\hline
$\gamma \gamma$ incl. & 0.77 $\pm$ 0.27 & (84\%,16\%) & \cite{CMS:2013comb,CMS:2013diphoton} \\
\hline
$\gamma \gamma$ dijet & 2.7 $\pm$ 1.9 & (23\%,77\%) & \cite{ATLAS:2012goa} \\
\hline
$\gamma \gamma$ dijet (tight) & 1.6 $\pm$ 0.9 & (24\%,76\%) & \cite{ATLAS:2013gg} \\
\hline
$\gamma \gamma$ dijet (loose) & 2.8 $\pm$ 1.8 & (45\%,55\%) & \cite{ATLAS:2013gg} \\
\hline
$\gamma \gamma$ dijet & 4.1 $\pm$ 2.4 & (27\%,73\%) & \cite{CMS:2013diphoton} \\
\hline
$\gamma \gamma$ dijet (tight) & 0.2 $\pm$ 0.8 & (21\%,79\%) & \cite{CMS:2013diphoton} \\
\hline
$\gamma \gamma$ dijet (loose) & 0.7 $\pm$ 1.15 & (47\%,53\%) & \cite{CMS:2013diphoton} \\
\hline
$ZZ^*$ & 1.7 $\pm$ 0.45 & (87\%,13\%) & \cite{ATLAS:2013ZZ} \\
\hline
$ZZ^*$ & 0.92 $\pm$ 0.28 & (87\%,13\%) & \cite{CMS:2013comb} \\
\hline
$WW^*$ & 1.0 $\pm$ 0.3 & (87\%,13\%) & \cite{ATLAS:2013comb} \\
\hline
$WW^*$ & 0.68 $\pm$ 0.20 & (87\%,13\%) & \cite{CMS:2013comb} \\
\hline
$b \bar{b}$ & -0.4 $\pm$ 1.0 & (0\%,100\%) & \cite{ATLAS:2013comb} \\
\hline
$b \bar{b}$ & 1.15 $\pm$ 0.62 & (0\%,100\%) & \cite{CMS:2013comb} \\
\hline
$b \bar{b}$ & 1.56 $\pm$ 0.72 & (0\%,100\%) & \cite{Knoepfel:2013qua} \\
\hline
$\tau \tau$ & 0.8 $\pm$ 0.7 & (80\%,20\%) & \cite{ATLAS:2013comb} \\
\hline
$\tau \tau$ (0/1 jet) & 0.77 $\pm$ 0.63 & (80\%,20\%) & \cite{CMS:2013comb} \\
\hline
$\tau \tau$ (VBF) & 1.40 $\pm$ 0.80 & (25\%,75\%) & \cite{CMS:2013comb} \\
\hline
$\tau \tau$ (comb.) & 1.10 $\pm$ 0.41 & (80\%,20\%) & \cite{CMS:2013comb} \\
\hline
\end{tabular}
\caption{Observed signal rates and production fractions which are employed in our fit.}
\label{HiggsSigStgns}
\end{table}
Rearranging the signal rates allows us to write them in terms of ratios of NP to SM cross sections, \ie
\begin{eqnarray}
&\displaystyle \frac{\sum_i \epsilon_i \sigma_i}{\sum_i \epsilon_i \sigma_i^{SM}} = \frac{\sum_i \epsilon_i (\sigma_i/\sigma_i^{SM}) (\sigma_i^{SM}/\sigma_{tot}^{SM})}{\sum_i \epsilon_i (\sigma_i^{SM}/\sigma_{tot}^{SM})} & \notag \\ \notag \\
& = \sum_i p_i^{SM} (\sigma_i/\sigma_i^{SM})&
\end{eqnarray}
where $p_i^{SM} = \epsilon_i (\sigma_i^{SM}/\sigma_{tot}^{SM})  [\sum_i \epsilon_i (\sigma_i^{SM}/\sigma_{tot}^{SM})]^{-1}$ is the SM probability for the $i^{th}$ production process. All NP effects are then parametrized by only two non-trivial ratios, 
\begin{eqnarray}
&\displaystyle \frac{\sigma_{ggF}}{\sigma_{ggF}^{SM}} = \frac{\Gamma_{gg}}{\Gamma_{gg}^{SM}} = \frac{\left| F_{1/2} ( \tau_t ) + c_\xi F_0 ( \tau_\xi ) \right|^2}{\left| F_{1/2} ( \tau_t ) \right|^2}& \notag \\
&\displaystyle \frac{\Gamma_{\gamma \gamma}}{\Gamma_{\gamma \gamma}^{SM}} = \frac{\left| F_1 ( \tau_W ) + \frac{4}{3} \left( F_{1/2} ( \tau_t ) + c_\xi F_0 ( \tau_\xi ) \right) \right|^2}{\left| F_1 ( \tau_W ) + \frac{4}{3} F_{1/2} ( \tau_t ) \right|^2} ,&
\end{eqnarray}
where $\tau_i$$=$$\ds 4 m_i^2 / m_h^2$, the loop functions, $F_{i}$, can be found in~\cite{Gunion:1989we}, and $c_\xi = 2 \ds (g_\xi / g_2) ( M_W^2 / m_\xi^2)$.

\begin{figure}[h!]
\includegraphics[scale=.625]{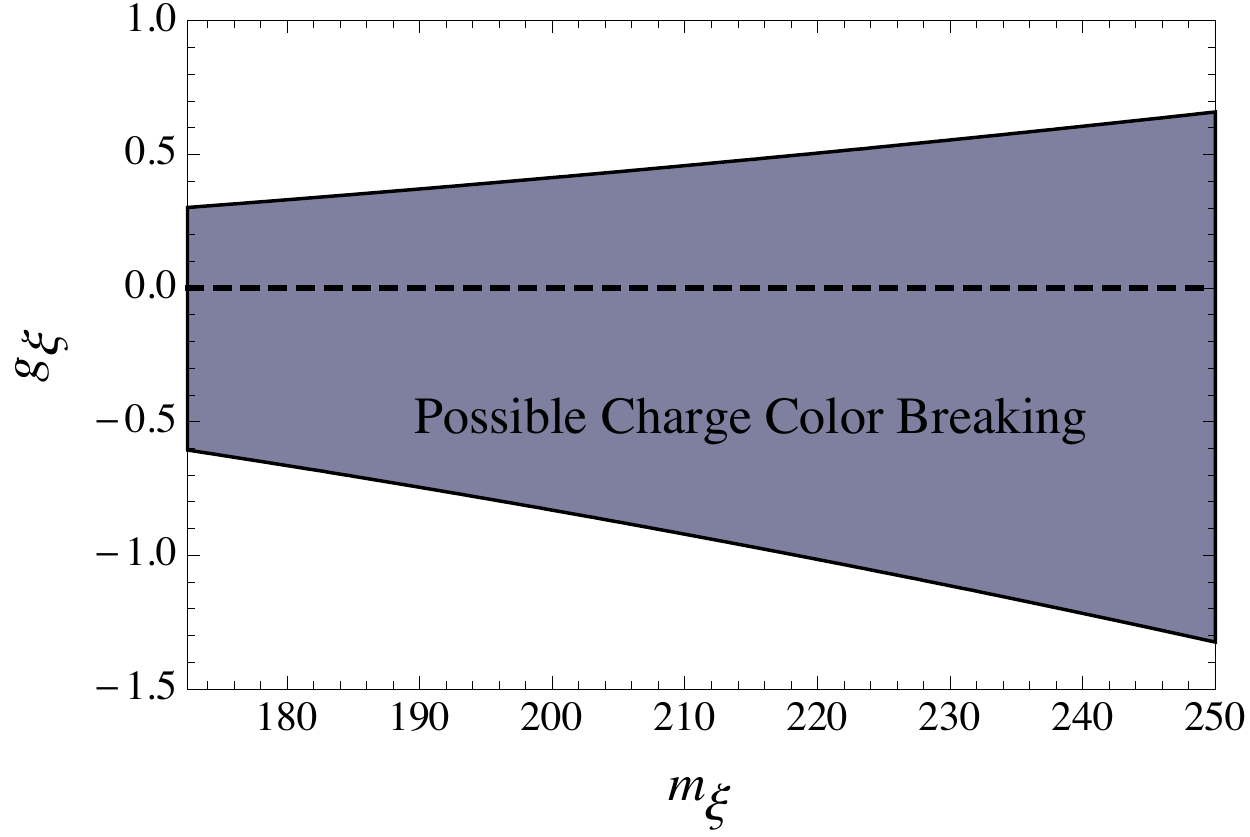}
\caption{\textit{The 68\% C.L. interval consistent with the current signal rates listed in TABLE~\ref{HiggsSigStgns}. Possible strong charge breaking minima become available if $g_\xi<0$.}}
\label{fig:HiggsFit}
\end{figure}

We perform a $\chi^2$ fit to the Higgs data in Table~\ref{HiggsSigStgns}, neglecting ttH production, and present the 68\% C.L. in FIG. \ref{fig:HiggsFit}. Good agreement with the data is achieved as long as $m_\xi$ $\geq$ $m_t$ and $g_\xi \lesssim 1/4$. For $g_\xi<0$, it is possible that color charge symmetry may be spontaneously broken. These scenarios have been analysed recently in detail~\cite{Patel:2013zla} and, although there may exist temperature regions in which a phase of spontaneous charge color breaking may be phenomenologically viable, we instead concentrate only on the zero temperature limit in which color symmetry is conserved. In this limit, we assume that $0<g_\xi\lesssim 1/4$, ensuring that the entire mass range of interest in FIG.~\ref{fig:BestFitRegions} is consistent with current Higgs data.

\subsection{Flavor Constraints}\label{sec:Flavor}

\begin{figure}[t!]
\includegraphics[scale=.75]{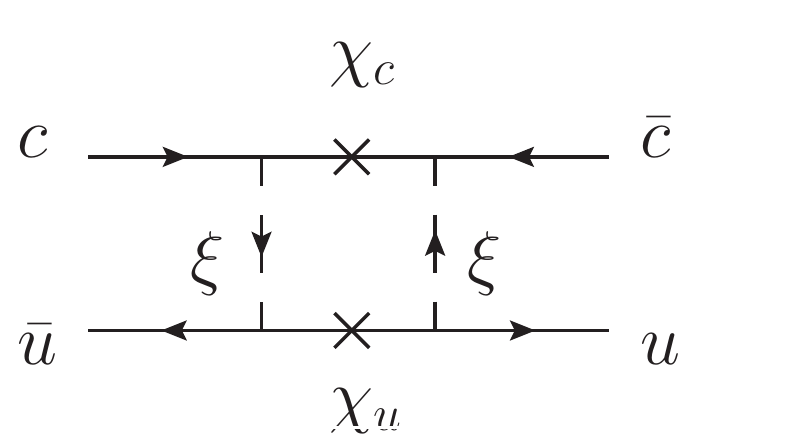}
\caption{\textit{New physics amplitudes for $D^0$-$\overline{D^0}$ mixing arising from scalar top partner exchange and Majorana mass insertions.}}
\label{fig:ddBarMixing_diagram}
\end{figure}
While our model avoids flavor constraints from the down-type quark sector, it is remains sensitive to $D^0$-$\overline{D^0}$ mixing through the process shown in FIG.~\ref{fig:ddBarMixing_diagram} due to Majorana mass insertions of $\chi_{u}$ and $\chi_c$. 

Mixing in the $D^0$ system arises from the off-diagonal matrix element of the mass matrix for the $D^0$ and $\overline{D^0}$ mesons, $(M - \frac{i}{2} \Gamma)_{12}$, with $M_{12}$ ($\Gamma_{12}$) associated with the dispersive (absorptive) part of the Hamiltonian. The relevant observable is the mass difference between the two eigenstates, which is related to the dispersive part of the Hamiltonian as $\Delta M_D = 2 | M_{12} |$ with 
\begin{eqnarray}
M_{12} = \frac{\left< \overline{D^0} \right| H_{eff}^{|\Delta C|=2} \left| D^0 \right>}{2 M_{D^0}} .
\end{eqnarray}

Although charm mixing experiences a significant GIM suppression in the SM, it is also plagued by large uncertainties, making it difficult to isolate NP effects. Nevertheless, we adopt the more conservative approach of Golowich {\it et al.}~\cite{Golowich:2007ka} and require that the NP contributions to $H_{eff}^{|\Delta C|=2}$ be within the 1$\sigma$ experimental limit. Furthermore, in the evaluation of the matrix elements, we use the numerical values from reference~\cite{Golowich:2007ka} for the decay constant, $f_{D^0}$, and non-perturbative parameter, $B_{D^0}$. 

The mass difference of the mass eigenstates is
\begin{eqnarray}
& \Delta M_D = \displaystyle \frac{\lambda_u^2 \lambda_c^2 f_{D^0}^2 M_{D^0} }{64 \pi^2 m_\xi^2}  B_{D^0} \left( 1 + \frac{2}{3} \frac{M_{D^0}^2}{( m_u + m_c)^2} \right) & \nonumber \\
& \times \beta( m_c , m_\xi ) | F ( x_{\chi_u} , x_{\chi_c} ) | .
\end{eqnarray}
We account for the QCD Renormalization Group (RG) running of the NP-generated operators from the NP scale, $m_\xi$, down to the charm mass with the factor,
\begin{eqnarray}
\beta( m_c , m_\xi ) = \left( \frac{\alpha_s ( m_\xi^2 )}{\alpha_s ( m_t^2 )} \right)^{1/7} \left( \frac{\alpha_s ( m_t^2 )}{\alpha_s ( m_b^2 )} \right)^{3/23} \left( \frac{\alpha_s ( m_b^2 )}{\alpha_s ( m_c^2 )} \right)^{3/25} \nonumber \\ 
\end{eqnarray}
where the running QCD coupling is evaluated at LO to match with NP results. The loop function, $F( x_{\chi_u}, x_{\chi_c} )$, which describes the short-distance NP contribution, is given by
\begin{eqnarray}
& F(x_{\chi_u}, x_{\chi_c} ) = \displaystyle \sqrt{x_{\chi_u} x_{\chi_u}} \Bigg( \frac{x_{\chi_c} \ln x_{\chi_c}}{(1 - x_{\chi_c} )^2 ( x_{\chi_u} - x_{\chi_c} )} & \nonumber \\
& \displaystyle + \frac{x_{\chi_c} - x_{\chi_u} ( 1 + x_{\chi_c} ) + x_{\chi_u}^2 - x_{\chi_u} ( 1 - x_{\chi_c} ) \ln x_{\chi_u}}{( 1 - x_{\chi_u} )^2 ( 1 - x_{\chi_c} ) ( x_{\chi_u} - x_{\chi_c} )} \Bigg)& \nonumber \\
\end{eqnarray}
with $x_{\chi_i} = m_{\chi_i}^2 / m_\xi^2$. \\

\begin{figure}[t]
\includegraphics[scale=.625]{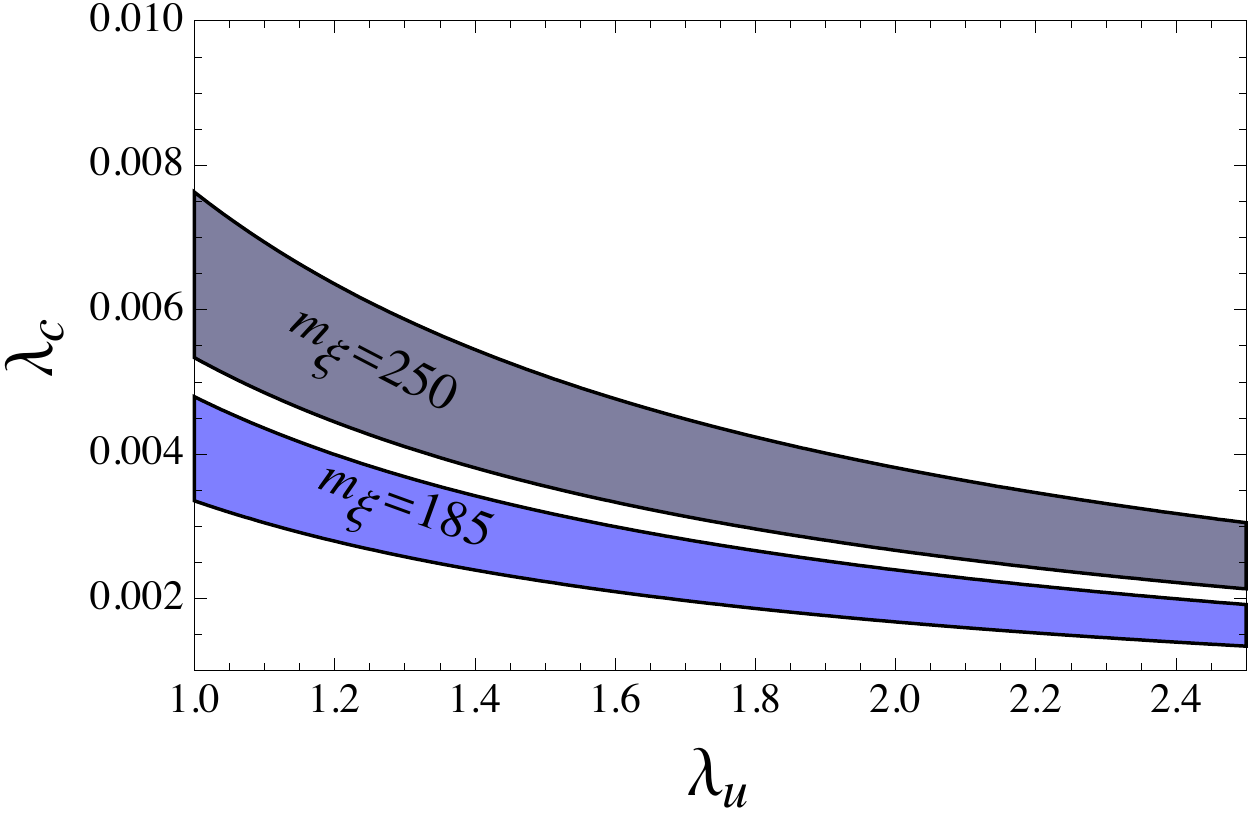}
\caption{\textit{1$\sigma$ allowed parameter regions for $\Delta M_D$ are shown for $m_\xi$ = 185 GeV (250 GeV) in light (dark) blue.}}
\label{fig:ddBarMixing_result}
\end{figure}

The 1$\sigma$ experimental bounds on  $\Delta M_D$~\cite{Beringer:1900zz}, shown in FIG.~\ref{fig:ddBarMixing_result}, indicate a weak dependence of $\Delta M_D$ on $m_{\xi}$. Furthermore, as agreement with the top data is good when $1 \lesssim \lambda_u \lesssim 2$ and 185 GeV $\lesssim m_\xi \lesssim$ 250 GeV, agreement with $\Delta M_D$ bounds require $\lambda_c \lesssim 10^{-3}$. Consequently, all collider effects from $\chi_{c}$ are highly suppressed and are henceforth treated as negligible\footnote{Although $\lambda_c$ can be re-generated at the loop level, it will necessarily be proportional to the mixing between the active and singlet neutrinos as well as the corresponding characteristic loop suppression, ensuring it stays within the bounds.}. 

\section{LHC Collider Signatures}\label{sec:LHC}

In addition to explaining the $A_{FB}^{t \bar{t}}$ and agreement with the top data from the Tevatron and the LHC, our model also predicts a number of collider signals that can be searched for at the LHC. First, single top quark production measurements as well as jets+$\met$ and monojet searches can potentially constrain the model. Moreover, the Majorana nature of the $\chi$ states can be probed in searches for like-sign top quark production. Finally, based on the global fit to the data in Section~\ref{sec:GlobalFit}, our model predicts a monotop signal with an associated lepton charge asymmetry. We discuss this signal in detail and outline strategies to search for it within the context of our model. In our analyses below, we explore all LHC signals in the ($m_\xi$, $\lambda_t$) region that is consistent with the top data (right panel of FIG.~~\ref{fig:BestFitRegions}), fixing $\lambda_u$ to its best-fit value.

We simulate all NP signals at the LHC for $\sqrt{s}=7~\tev$ using MadGraph5v1.5.10~\cite{Alwall:2011uj} interfaced with PYTHIA~\cite{Sjostrand:2006za,Sjostrand:2007gs} to account for parton shower and hadronization effects. The 3-body $\chi_u$ decays are complicated to simulate as they involve displaced vertices and highly non-trivial $\met$ reconstruction. We postpone a more realistic treatment of these effects for future work, instead presenting our results at the parton level only. We further assume the narrow width approximation for $\chi_u$ decays, convolving the production cross section with the appropriate branching ratio. A proper treatment of these effects will further reduce our NP signals, consequently, our results should be seen as the most optimistic scenario. 

\subsection{Single top quark production}\label{sec:singletop}

\begin{figure}[t!]
\includegraphics[scale=.5]{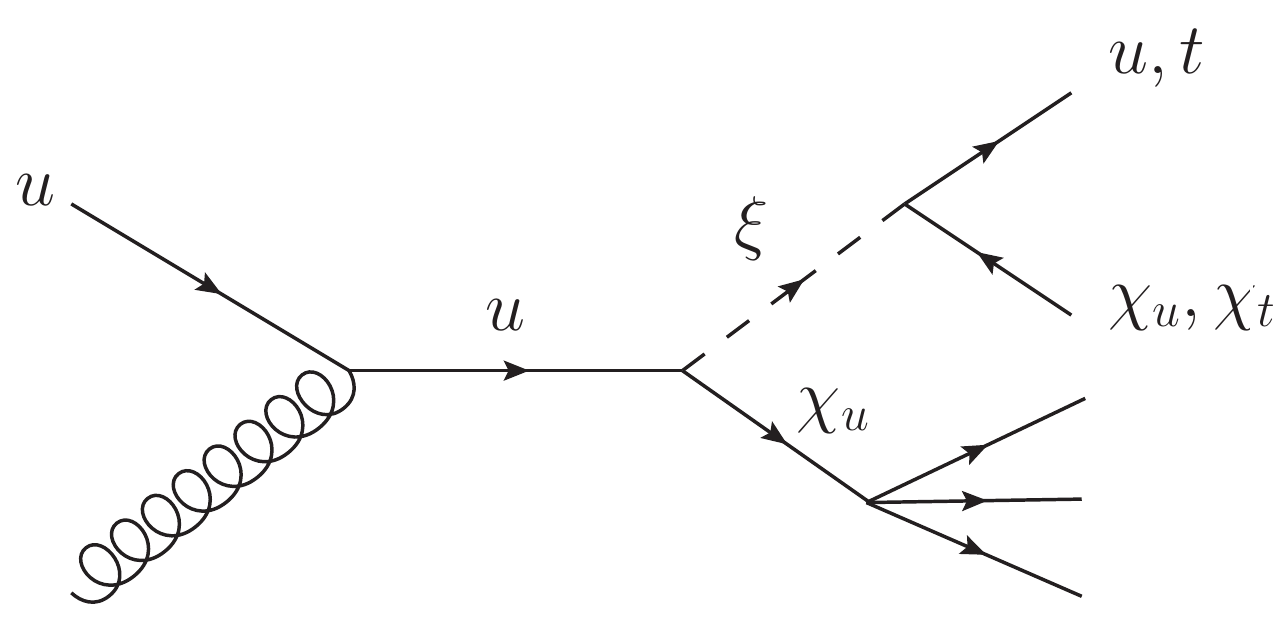}
\caption{\textit{The dominant NP process contributing to the NP signatures we consider at the LHC. Depending on the decay mode of $\chi_u$, this process can contribute to single top production or generate a monojet or monotop signal.}}
\label{fig:MonotopFig}
\end{figure}

The ATLAS collaboration has measured the cross section for single top quark production in the $t$-channel, $\sigma ( t \bar{b} j ) = 59^{+18}_{-16}$pb~\cite{Aad:2012ux}, while the SM prediction is known up to NNLO and is $\sigma (t \bar{b} j ) = 41.9^{+1.8}_{-0.8}$pb~\cite{Kidonakis:2011wy}. The dominant contribution to this process in our model is through quark-gluon fusion, $ug \to t \chi_u \chi_t + h.c.$ (FIG.~\ref{fig:MonotopFig}), where $\chi_u$ decays through the neutral current-current interaction to jets + $\met$($\nu$). Assuming a b-tagging efficiency of $\epsilon_b$=$0.5$, corresponding to a mis-identification rate $\slashed{\epsilon}_{c \to b}$=$0.08$ ($\slashed{\epsilon}_{j \to b}$=$0.002$) for charm (light) jets~\cite{Aad:2009wy}, we estimate the total NP contribution to single top production as 
\begin{eqnarray}
& \sigma_{t \bar{b} j}^{NP} = \sigma_{t \chi_u \chi_t} \; 2 \bigg( \textrm{BR} ( \chi_u \to \nu b \bar{b} ) \epsilon_b (1-\epsilon_b) & \nonumber \\
& + \textrm{BR} ( \chi_u \to \nu c \bar{c} ) \slashed{\epsilon}_{c \to b} + \textrm{BR} ( \chi_u \to \nu j j ) \slashed{\epsilon}_{j \to b} \bigg) & \nonumber \\
& = 0.03 \; \sigma_{t \chi_u \chi_t} \; .
\end{eqnarray} 
This cross section lies within the theoretical error of the SM prediction for the entire parameter region of interest as shown in FIG.~\ref{fig:constraints}, i.e., the model remains unconstrained from single top production measurements.

\subsection{Jets+$\mathbf \met$ production}\label{sec:jets_met}

On-shell production of $\xi\xi^{*}$ with subsequent decays to $u \bar{u} \chi_u \chi_u$ with $\chi_u \to \met (3 \nu)$ generates a jets+$\met$ signature at the LHC (FIG.~\ref{fig:AFB_process}). These invisible $\chi_u$ decays provide significant branching ratio suppression, i.e., $\sigma_{jets+\met} = \sigma_{u \bar{u} \chi_u \chi_u} \textrm{BR}( \chi \to 3 \nu )^2 = 0.36\%  \; \sigma_{u \bar{u} \chi_u \chi_u}$. 

Current jets+$\met$ searches by ATLAS and CMS place stringent constraints on the NP parameter space. However, such searches are typically geared towards heavy NP, implementing high $\met$ cuts. The most relevant search for our model is from signal region A in the ATLAS search~\cite{ATLAS:2011met}, which requires at least 2 jets and the following cuts:
\begin{eqnarray}
& \met > 130 \; \textrm{GeV} \qquad p_T^{j1,j2} > (130, 40) \; \textrm{GeV} \nonumber & \\
&  m_{eff} > 1 \; \textrm{TeV} & 
\end{eqnarray}
where $p_T^{j1,j2}$ is the $p_T$ of the first and second hardest jet while $m_{eff} \equiv \met + \sum_i |p_T^i|$. The result of this search is the upper bound $\sigma < 22$ fb, after including acceptance and efficiency~\cite{ATLAS:2011met}. 

As is shown in FIG.~\ref{fig:constraints}, the parton level cross section (before cuts) in our model is larger than the ATLAS bound in some regions of parameter space by an $\mathcal{O}$(1) factor. However, as the signal originates from on-shell $\xi \xi^*$ production, which only allows for $m_{eff} \lesssim 500$ GeV, the acceptances are negligible ($\lesssim$ 1\%). In addition, a more realistic treatment of the $\met$ reconstruction in $\chi_u$ decays will only further reduce the acceptance. For these reasons, our signal remains consistent with current jets+$\met$ searches. However, given the proximity of the predicted production cross section to the existing bound, we feel that it would be worthwhile for future LHC searches to be geared towards probing light new particles as well by relaxing kinematic cuts on $\met$ and $m_{eff}$.

\subsection{Monojet production}\label{sec:monojets}

In our model, a monojet signal arises in quark-gluon fusion, $u g \to u \chi_u \chi_u + h.c.$ (FIG.~\ref{fig:MonotopFig}), when both $\chi_u$'s decay invisibly. While the model of ref.~\cite{Isidori:2011dp} is highly constrained by monojet searches, our modified setup evades such bounds by accounting for the $\chi_u \to 3 \nu$ branching ratio suppression, yielding a production cross section $\sigma_{monojet} = \sigma_{u \chi_u \chi_u} \textrm{BR}( \chi \to 3 \nu )^2 = 0.36\% \; \sigma_{u \chi_u \chi_u}$. 
\begin{figure*}[t!]
\centering
\mbox{
\subfigure{
\includegraphics*[scale=.625]{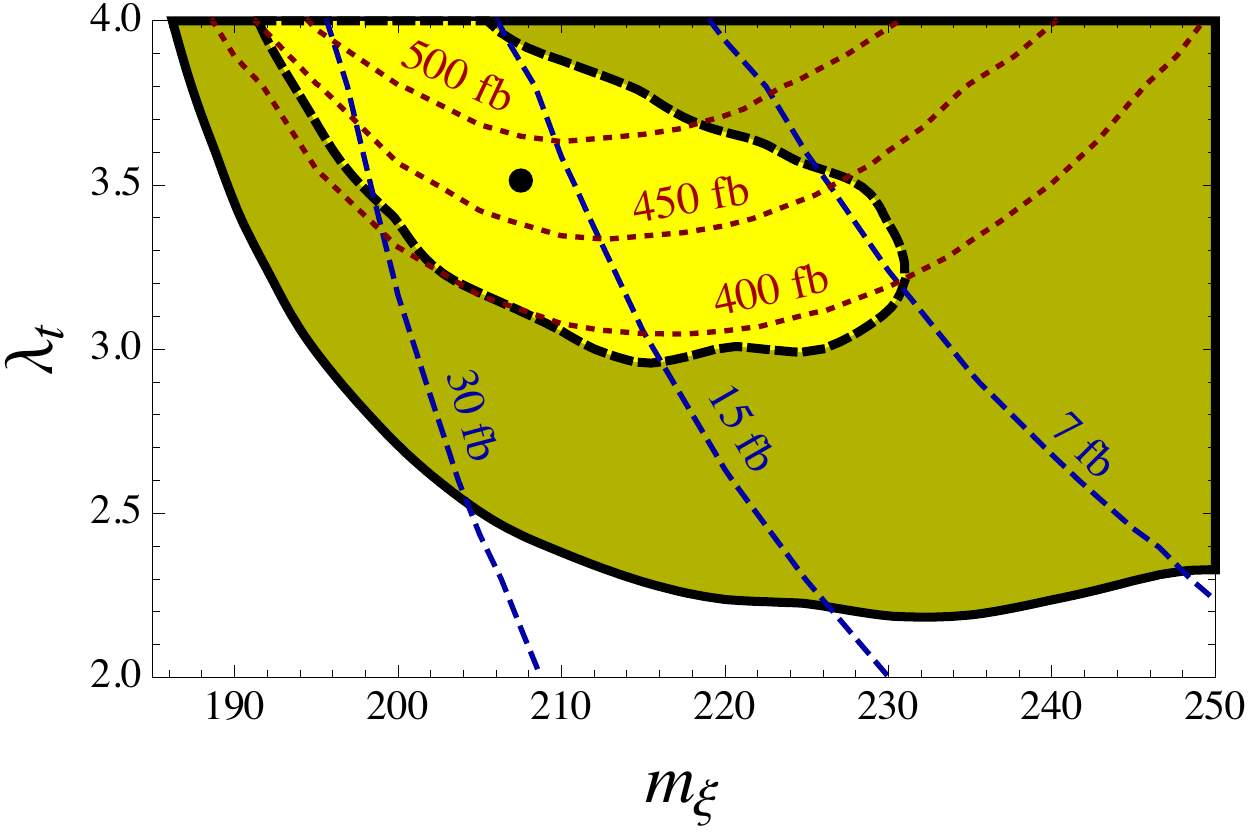}
}
\quad \quad \quad \subfigure{
\includegraphics*[scale=.625]{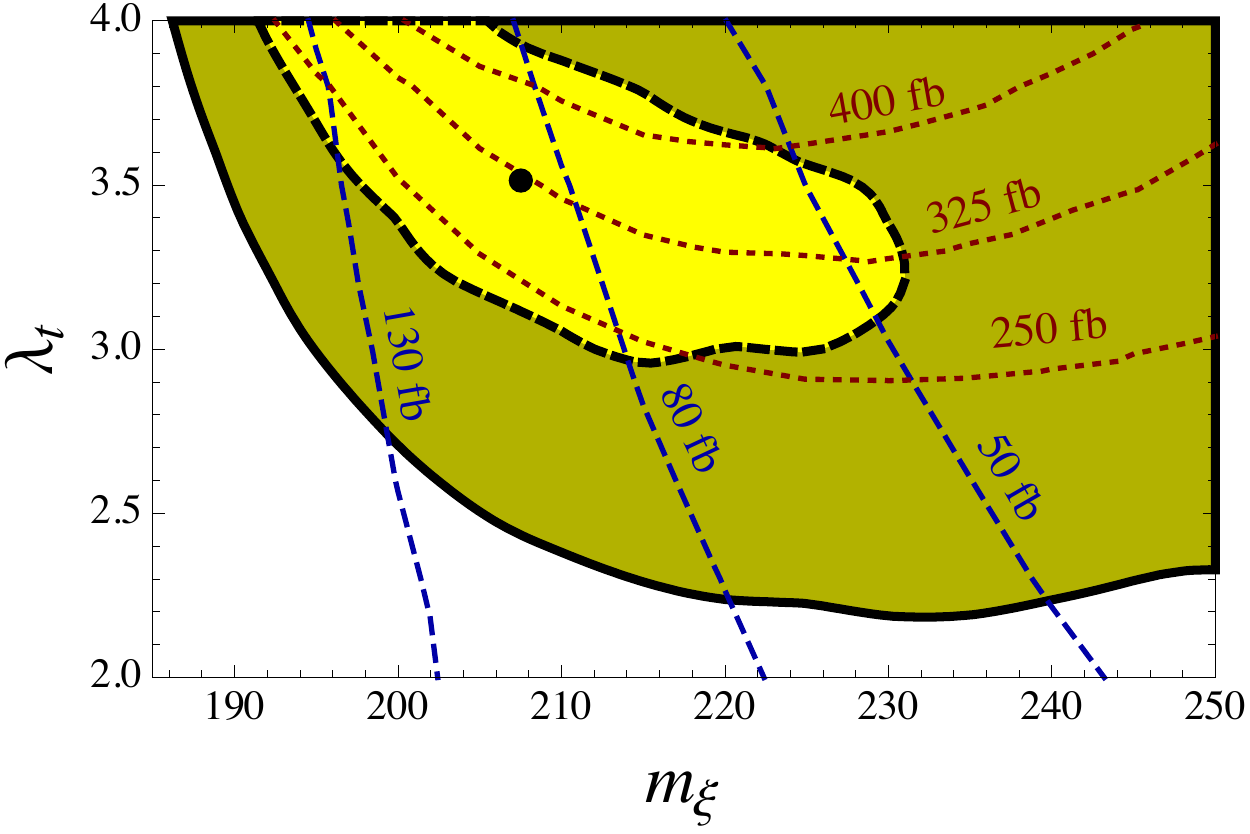}
} 
}
\caption{{\it All results are superimposed on top of the best fit regions from FIG.~\ref{fig:BestFitRegions}.} Left panel: {\it Predictions for the NP single top (jets+$\met$) production cross section are shown as red dots (blue dashes). The NP single top cross section lies within the theoretical error throughout the entire parameter region of interest. Branching ratio suppression from $\chi_u$ decays is not sufficient to fully evade bounds from jets+$\met$ searches, however, the acceptance rate for $m_{eff}$ is negligible. Large sections of the preferred parameter space can potentially be ruled out if cuts on $m_{eff}$ are relaxed.} Right panel: {\it Predictions for the monojet (same-sign top) production cross section are shown in blue dashes (red dots). Branching ratio suppression from $\chi_u$ decays is sufficient to evade the existing upper bound of 1.7pb on monojet production. Accounting for an acceptance of $\sim$25\% from the cut on the invariant mass of the muon pair yields a result that is below the current bound of 3.7pb for same-sign top production.}}
\label{fig:constraints}
\end{figure*}

The ATLAS collaboration has performed such a search~\cite{ATLAS:2011monojet}, separating events into 3 categories based on the $p_T$ of the jet. The search relevant to our model is characterized by the LowPT cut of $p_T > 120$ GeV, which yields a model-independent 95\% C.L. upper limit on the production cross section times acceptance of 1.7 pb. CMS has also performed monojet searches~\cite{CMS:2012monojet}, placing similar limits.

Our prediction for the parton level monojet production cross section, shown in the right panel of FIG.~\ref{fig:constraints}, is well within the existing bounds throughout the parameter region of interest.

\begin{figure*}[t!]
\centering
\mbox{
\subfigure{
\includegraphics*[scale=.625]{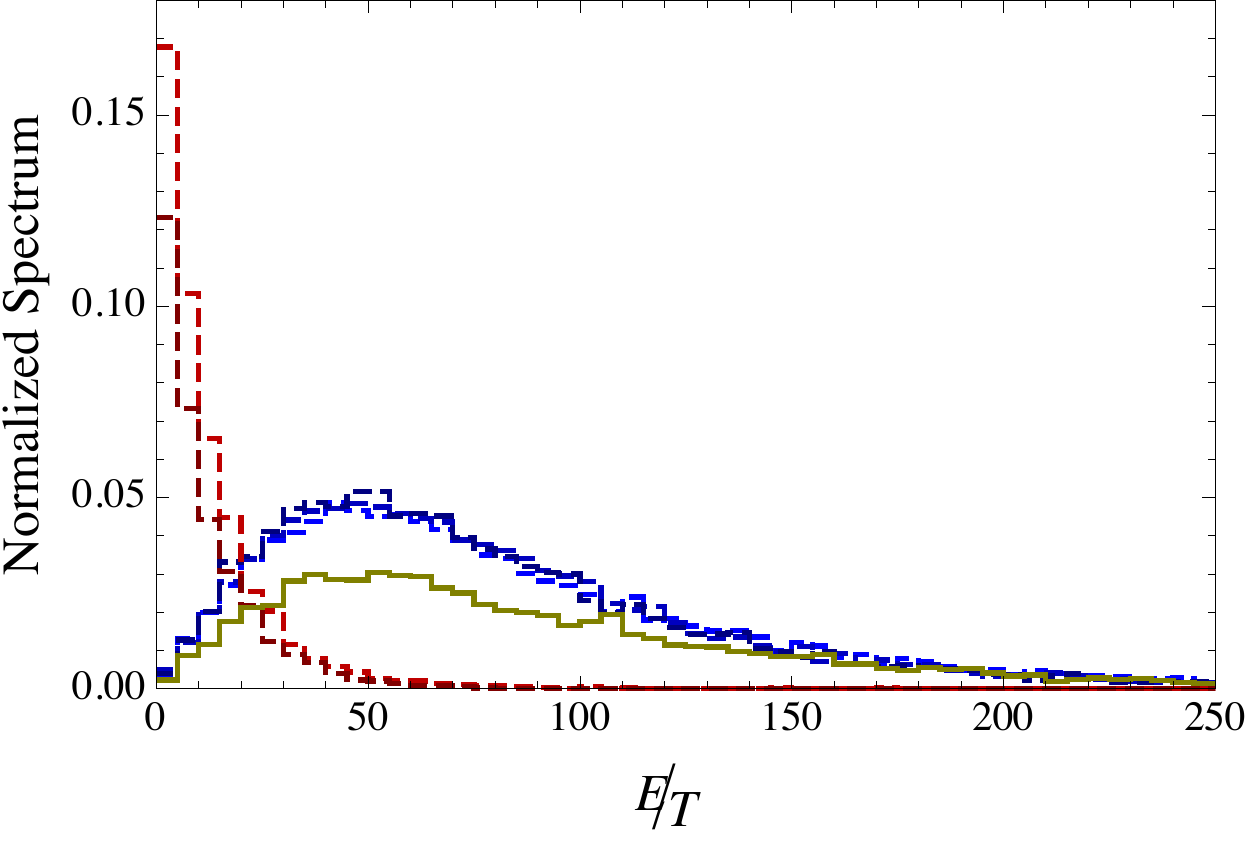}
}
\quad \quad \quad \subfigure{
\includegraphics*[scale=.625]{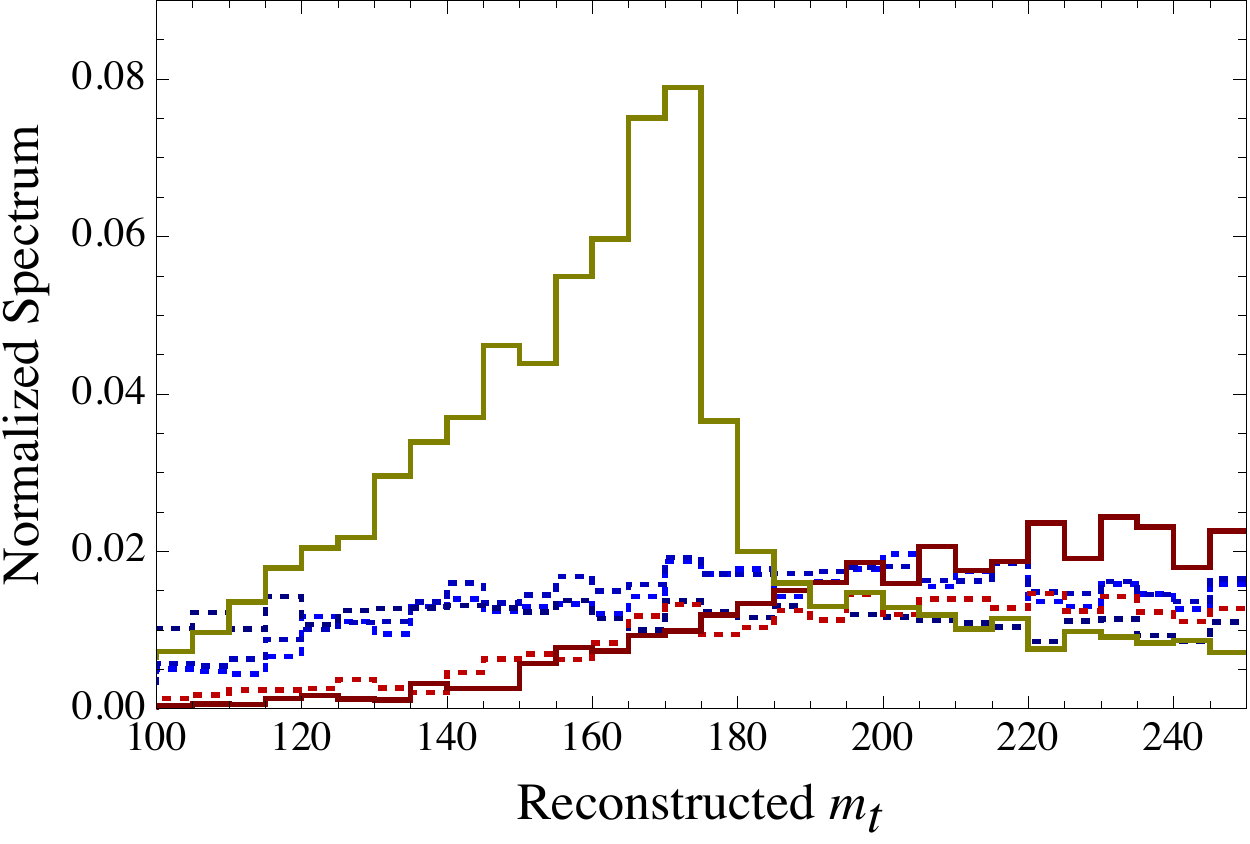}
} 
}
\caption{{\it In both panels, the backgrounds are presented as dashed lines with the $jjjZ$ and $bbjZ$ backgrounds shown in blue while the single and pair top related backgrounds are shown in red. The signal distribution, in solid gold, is calculated at the best fit point.} Left panel: {\it The normalized spectra of $\met$ at the LHC for $\sqrt{s}$=7 TeV. Due to the production process, both $\chi_u$ and $\chi_t$ are highly boosted along the beam direction, suppressing the $\met$ and making it difficult to distinguish signal from background.} Right panel: {\it The normalized spectra of the reconstructed top mass at the LHC for $\sqrt{s}$=7 TeV. All backgrounds are stacked with the solid red line representing the total background. The $jjjZ$ and $bbjZ$ backgrounds contain no top quark and are therefore flat throughout their spectra while the top related backgrounds are highly suppressed by a $\met$ cut of $\met >$50 GeV. The signal maintains a clear peak near the top mass, distinguishing it from the backgrounds in the region 120 GeV $< m_{t,r} <$ 180 GeV.}}
\label{fig:distr_plots}
\end{figure*}

\subsection{Same-sign top quark production}\label{sec:samesigntop}

The Majorana nature of $\chi_u$ can be directly probed by searches for same-sign top quark production. A $tt$ final state is generated in a process similar to that in FIG.~\ref{fig:AFB_process}, i.e., from two initial state $u$ quarks with a $\chi_u$ mass insertion in the $t$-channel. In addition, this process generates a same-sign dilepton signal when the same-sign top quarks decay semi-leptonically.

The ATLAS collaboration has performed a search for anomalous production of two muons with the same electric charge, obtaining the best expected limits when the invariant mass of the muon pair is $\geq 200$ GeV. Using these results, same-sign top production is also constrained, deriving a model-independent 95\% C.L. upper limit on the production cross section of 3.7pb~\cite{Aad:2012cg}. The CMS collaboration has also performed a search for same-sign top quark production~\cite{Chatrchyan:2011dk}, however, their model-independent limit on the production cross section, 17 pb, is weaker than the above ATLAS bound.

In our analysis, we simulate same-sign top quark production, allowing PYTHIA to decay the top quarks and extracting only events which correspond to semi-leptonic decays to muons. Applying the dimuon invariant mass cut from the ATLAS search, we find an overall acceptance rate of $\sim$25\% that is largely independent of $\lambda_u$,$\lambda_t$, and $m_\xi$. The resulting prediction for the same-sign top quark production cross section is well below the current limit and is shown in FIG.~\ref{fig:constraints}. 


\subsection{Monotop production}\label{sec:monotop}

Finally, we discuss the monotop signal in our model, which is generated by quark-gluon fusion, $u g \to t \chi_u \chi_t + h.c.$ (FIG.~\ref{fig:MonotopFig}), with $\chi_u$ decaying invisibly. The invisible $\chi_u$ decays suppress the signal production cross section by one factor of the $\chi_u \to 3 \nu$ branching ratio, yielding $\sigma_{t + \met} = 6\% \; \sigma_{t \chi_t \chi_u}$.

In determining a viable search strategy at the LHC, we focus solely on the hadronic decay of the top quark, yielding the signal $ p p \to t + \met \to b W + \met \to b j j + \met$ with $b$ a b-tagged jet and $j$ a light quark or gluon jet. The dominant SM backgrounds are due to $p p \to j j j Z \to j j j \nu \nu$ with a light jet mis-identified as a $b$ jet, $p p \to b \bar{b} j Z \to b \bar{b} j \nu \nu$ with a $b$-jet not tagged, as well as backgrounds from both single and pair production of top quarks with some jets not detected. Again, we choose a b-tagging efficiency of $\epsilon_b$=$50\%$, corresponding to a mis-identification rate of $\slashed{\epsilon}_{c \to b}$=$8\%$ ($\slashed{\epsilon}_{j \to b}$=$0.2\%$) for charm (light) jets~\cite{Aad:2009wy}. The monotop production cross section, $\sigma_{t + \met} \sim \mathcal{O}$(1~pb), is swamped by the combined effect of the SM backgrounds, $\sim \mathcal{O}$(100 pb). However, by implementing a specific set of kinematic cuts, a statistically significant signal can be distinguished from the background. 

We begin by considering the normalized $\met$ distributions, shown in the left panel of FIG.~\ref{fig:distr_plots}. In many NP models, the signal is typically produced by a heavy resonance (much heavier than $m_t$) that decays on-shell to $t + \met$, generally focusing the $\met$ distribution in the $>$100 GeV region~\cite{Andrea:2011ws,Wang:2011uxa}. In contrast to this, the background $\met$ distributions are concentrated in the $<$100 GeV. Hence, in these cases, a $\met$ cut is a sufficient discriminator. 

In our model, the $\met$ is the vector sum of the $p_T$ of $\chi_u$ (as $\chi_u \to 3 \nu$) and $\chi_t$. $\chi_u$ is produced in association with $\xi$ via a highly off-shell $s$-channel $u$ quark, leading to a $\chi_u$-$\xi$ system which is highly boosted along the beam direction. As $\chi_t$ is produced nearly at rest in the $\xi$ rest frame (due to the small mass splitting $m_\xi$-$m_t$), $\chi_t$ inherits a large fraction of the $\xi$ boost, leading to a highly collimated $\chi_u$-$\chi_t$ state. Furthermore, the large boost along the beam direction limits the $p_T$ of both $\chi_u$ and $\chi_t$ individually, concentrating the signal $\met$ in the $<$100 GeV region, as shown in the left panel of FIG.~\ref{fig:distr_plots}. As a result, the monotop signal is not easily distinguished from the background by a $\met$ cut alone and further cuts are necessary. Nevertheless, we apply a cut on the $\met$ such that we only retain events for which $\met>$ 50 GeV, effectively suppressing the bulk of all top-related backgrounds (see the left panel of FIG.~\ref{fig:distr_plots}). 

\begin{figure}[t!]
\includegraphics*[scale=.625]{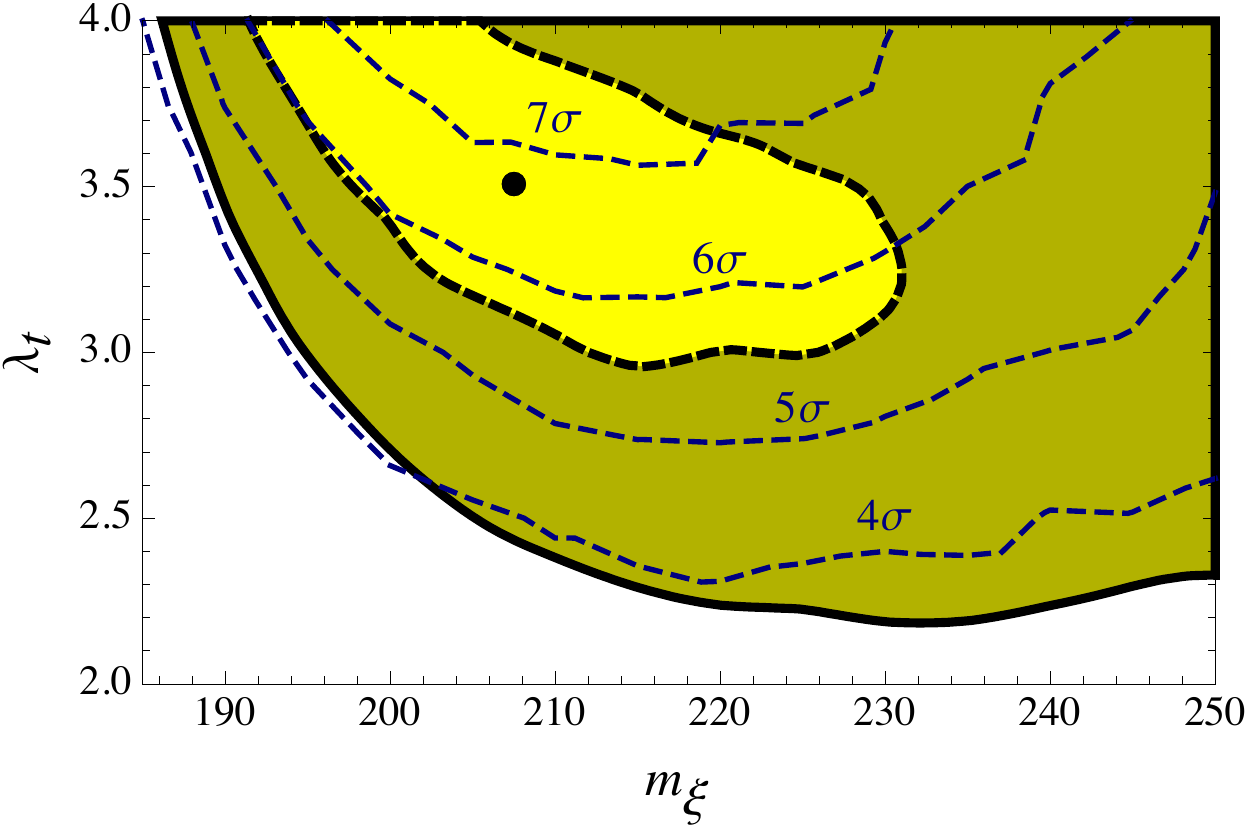}
\caption{The discovery potential of our monotop signal for 1 fb$^{-1}$ of LHC data at $\sqrt{s}$ = 7 TeV. Contours of significance for the signal-to-background ratio are superimposed on the best fit regions in FIG.~\ref{fig:BestFitRegions}. }
\label{fig:S2B}
\end{figure}

Next, we reconstruct the top quark mass by calculating the invariant mass of the leading $b$-jet along with the two leading light jets. The top related backgrounds are highly suppressed by the $\met$ cut discussed above and the remaining $jjjZ$ and $b\bar{b}jZ$ backgrounds yield flat distributions as they do not contain a top quark. There is a clear peak from the signal above the background in the region 120 GeV $\lesssim m_{t,r} \lesssim$ 180 GeV, as shown in the right panel of FIG.~\ref{fig:distr_plots}. We calculate the signal-to-background ratio, $S / \sqrt{B}$, in this kinematic region for the entire ($\lambda_t$, $m_\xi$) parameter space of interest, fixing $\lambda_u$ to its best fit value. We present the results as contours of significance in the $\lambda_t$-$m_\xi$ plane (FIG.~\ref{fig:S2B}) which represent the discovery potential for 1 fb$^{-1}$ of LHC data at $\sqrt{s}$ = 7 TeV. The contours are superimposed over the best fit regions of FIG.~\ref{fig:BestFitRegions} to demonstrate that, if our model is responsible for the $A_{FB}^{t \bar{t}}$, there exists a high chance for observation of a monotop signal at the LHC. We further note that our results can be easily scaled to incorporate the total amount of data collected during the 7 TeV run at the LHC.

\subsection{Lepton charge asymmetry}\label{sec:lep_CA}

The monotop signal is further distinguished by the existence of an associated lepton charge asymmetry, which measures the net charge difference in the semi-leptonic top decays. At the LHC, our monotop signal produces many more tops than anti-tops, as the CP-conjugate process has a much smaller cross section due to pdf suppression of the $\bar{u}$ vs. $u$ quark. This asymmetry is defined as $A_C^\ell = (N_{\ell^+} - N_{\ell^-}) / (N_{\ell^+} + N_{\ell^-})$, where $N_{\ell^\pm}$ is the number of monotop events with a single positively/negatively charged lepton. This definition can also be factorized into the SM and NP contributions as in eq.~(\ref{AFB_factorization}), i.e., 
\begin{eqnarray}
 A_C^\ell =  A_C^{\ell, SM} \left( 1 + \displaystyle \frac{\sigma^{NP}_{t+\met}}{\sigma^{SM}_{t + \met}} \right)^{-1}  + A_C^{\ell, NP} \left( 1 + \displaystyle \frac{\sigma^{SM}_{t + \met}}{\sigma^{NP}_{t + \met}} \right)^{-1} . \nonumber \\
 \label{eq:lepton_charge_asymmetry}
\end{eqnarray} 
As the monotop production cross section in the SM is both loop and CKM suppressed, the NP cross section dominates, i.e., $\sigma^{SM}_{t+\met} / \sigma^{NP}_{t+\met} << 1$. This highly damps the effects of the SM asymmetry in eq.~(\ref{eq:lepton_charge_asymmetry}), simultaneously enhancing the NP asymmetry such that $A_C^\ell \approx A_C^{\ell, NP}$. This type of asymmetry has also been studied in single top production, e.g., see~\cite{Bowen:2005xq,Craig:2011an} and references within. However, in the single top scenario the situation is reversed, i.e., $\sigma^{NP}_{tbj} / \sigma^{SM}_{tbj} <<1$, instead favoring the SM asymmetry over the NP contribution. Therefore, for our model, the effects of NP in this asymmetry may only be easily deduced in the monotop production channel, which has yet to be observed. 

The NP asymmetry is independent of the couplings as they cancel in the ratio of forward and backward cross sections. In addition to this, the asymmetric part of the cross section has very little $m_\xi$ dependence, leading to an essentially parameter independent prediction for the asymmetry of $A_C^\ell \approx 80 \%$. If a monotop signal is observed at the LHC, an associated search for the presence (or absence) of this asymmetry will serve to further identify its phenomenological origin.


\section{Conclusions}
\label{sec:Conclusions}

We have shown that the anomalous top quark forward-backward asymmetry, $A_{FB}^{t \bar{t}}$, may well originate from on-shell production and decay of scalar top partners to a $t \bar{t} + \met$ final state. This class of models is interesting as, although they generate $A_{FB}^{t \bar{t}}$ through $t$-channel exchange of light mediators, they avoid strong atomic parity violation constraints by forgoing a direct interaction between the first generation quarks and the top quark. Moreover, constraints from Higgs searches and flavor observables are quite mild. However, jets+$\met$ and monojet searches at the LHC strongly constrain these models. 


We have introduced a model containing a triplet of Majorana fermions that can be interpreted as right-handed sterile neutrinos. Mixing with the active neutrino sector via a Type-I seesaw operator opens up new decay channels for these fermions which play a crucial role in evading existing LHC bounds, particularly from jets+$\met$ and monojet searches. However, future LHC searches could well probe our model in its entirety. 

The model also predicts a monotop signal that can be efficiently distinguished from the SM background by a combination of cuts on $\met$ and the reconstructed top quark mass. We have explored the discovery potential for this signal at the LHC in the same region of parameter space that favors the $A_{FB}^{t \bar{t}}$ and conclude that there exists a high possibility for observation. Moreover, we have also identified a distinctive lepton charge asymmetry associated with monotop production in our model. If monotops are observed at the LHC, performing an associated search for this asymmetry will help in resolving the phenomenological origin of such a signature.

\acknowledgements We thank J.~Kamenik, T.~Martin and S.~Tulin for many insightful and useful discussions. This work was supported in part by a Natural Sciences and Engineering Research Council (NSERC) of Canada Discovery Grant. A.K., J.N. and P.W. are supported by NSERC.

\end{document}